\documentstyle[a4p,12pt,epsfig,subfigure,array,cite,pennames]{article}
\def\Zzero{Z$^{0}$}

\newcommand{\inmath}[1] {\ifmmode#1\else$#1$\fi}
\newcommand{\definmath}[2] {\def#1{\ifmmode#2\else$#2$\fi}}
\def\qq{\ifmmode {{\mathrm q\bar{\mathrm q}}}
    \else {${\mathrm q\bar{\mathrm q}}$} \fi}
\definmath{\Pq}      {\mathrm{q}}
\definmath{\Paq}  {\overline{\mathrm{q}}}
\def\K{\ifmmode {{\mathrm K}} \else {${\mathrm K}$} \fi}
\def\Zzero{\ifmmode {{\mathrm Z}^0} \else {${\mathrm Z}^0$}\fi}
\def\dEdx{\ifmmode {{\mathrm d}E/{\mathrm d}x}\else
                   {${\mathrm d}E/{\mathrm d}x$}\fi }
\def\dedx{\ifmmode {{\mathrm d}E/{\mathrm d}x}\else
                   {${\mathrm d}E/{\mathrm d}x$}\fi }
\newcommand{\qqbar}  {\Pq\Paq}

\def\costhstar{\ifmmode {\cos\theta} \else $\cos\theta$ \fi }

\newcommand{\rh}[2]{\mbox{$\rho_{{#1}{#2}}$}}
\renewcommand{\Re}{\mbox{\rm Re\,}}
\renewcommand{\Im}{\mbox{\rm Im\,}}

\newcommand{\CTH}{\mbox{$\cos\theta_{\mathrm H}$}}

\newcommand{\PHIH}{\mbox{$\phi_{\mathrm H}$}}

\newcommand{\ks} {$\mathrm K^{*0}$}
\newcommand{\KS} {$\mathrm K^*(892)^0$}

\newcommand{\Kpi}{K$^\pm\pi^\mp$}
\newcommand{\likesign}{K$^\pm\pi^\pm$}
\newcommand{\pipi}{$\pi^+\pi^-$}
\newcommand{\rzz}{$\rho(770)^0$}
\newcommand{\rz}{$\rho^0$}
\newcommand{\omm}{$\omega(782)$}
\newcommand{\om}{$\omega$}
\newcommand{\Rzz}{\rh{0}{0}}
\newcommand{\Rer}{\Re \rh{1}{-1}}
\newcommand{\xp}{$x_p$}
\newcommand{\frag}{$(1/\sigma_{\mathrm h}){\mathrm d}\sigma/{\mathrm d}x_p$}

\newcommand{\ddiffc}{$(1/\sigma_{\mathrm h})
{\mathrm d}^2\sigma/{\mathrm d}{x_p}{\mathrm d}\cos \theta_{\mathrm H}$}
\newcommand{\ddiffa}{$(1/\sigma_{\mathrm h})
{\mathrm d}^2\sigma/{\mathrm d}{x_p}{\mathrm d}|\alpha|$}
\newcommand{\eetoZtoqq}{e$^+$e$^-\rightarrow{\mathrm Z}^0
\rightarrow{\mathrm q}{\bar {\mathrm q}}$}

\begin{document}

\begin{titlepage}
\begin{center}{\large   EUROPEAN LABORATORY FOR PARTICLE PHYSICS
}\end{center}\bigskip
\begin{flushright}
       CERN-PPE/97-094  \\ 24th July 1997 
\end{flushright}
\bigskip\bigskip\bigskip
\begin{center}
    \huge\bf\boldmath
Spin alignment of leading \KS\ mesons in hadronic Z$^0$ decays 
\end{center}\bigskip\bigskip

\begin{center}
{\LARGE The OPAL Collaboration}
\end{center}\bigskip\bigskip
\bigskip\begin{center}{\large  Abstract}\end{center}
\noindent 
Helicity density matrix elements for 
inclusive \KS\ mesons from hadronic Z$^0$ decays
have been measured over the full range of \ks\ momentum 
using data taken with the OPAL experiment at LEP. 
A preference for occupation of the helicity zero state is 
observed at all scaled momentum \xp\ values above 0.3, with the matrix 
element \rh{0}{0} rising to
$0.66 \pm 0.11$ for \xp$>0.7$. The values of the real part of the 
off-diagonal element \rh{1}{-1} are negative at large \xp, with a
weighted average value of $-0.09\pm0.03$ for \xp$>0.3$, in 
agreement with new theoretical predictions based on Standard Model
parameters and coherent fragmentation of the \qqbar\ system from
the Z$^0$ decay. 
All other helicity density matrix elements
measured are consistent with zero over the entire \xp\ range. 
The \ks\ fragmentation function has also been measured and the total
rate determined to be $0.74 \pm 0.02 \pm 0.02$ \KS\ mesons per
hadronic Z$^0$ decay.

\bigskip\bigskip\bigskip\bigskip
\bigskip\bigskip\bigskip
\bigskip
\bigskip
\begin{center}
% {\large Submitted to Zeitschrift f\"ur Physik C}
{\large To be submitted to Physics Letters B}
\end{center}
%\vfill
\end{titlepage}

\begin{center}{\Large        The OPAL Collaboration
}\end{center}\bigskip
\begin{center}{
%begin authorlist
K.\thinspace Ackerstaff$^{  8}$,
G.\thinspace Alexander$^{ 23}$,
J.\thinspace Allison$^{ 16}$,
N.\thinspace Altekamp$^{  5}$,
K.J.\thinspace Anderson$^{  9}$,
S.\thinspace Anderson$^{ 12}$,
S.\thinspace Arcelli$^{  2}$,
S.\thinspace Asai$^{ 24}$,
D.\thinspace Axen$^{ 29}$,
G.\thinspace Azuelos$^{ 18,  a}$
% ...
%end authorlist
}\end{center}\bigskip
\bigskip
%begin institutes
$^{  1}$School of Physics and Space Research, University of Birmingham,
Birmingham B15 2TT, UK
\newline
$^{  2}$Dipartimento di Fisica dell' Universit\`a di Bologna and INFN,
I-40126 Bologna, Italy
\newline
% ...
$^{  5}$Cavendish Laboratory, Cambridge CB3 0HE, UK
\newline
% ...
$^{  8}$CERN, European Organisation for Particle Physics,
CH-1211 Geneva 23, Switzerland
\newline
$^{  9}$Enrico Fermi Institute and Department of Physics,
University of Chicago, Chicago IL 60637, USA
\newline
% ...
$^{ 12}$Indiana University, Department of Physics,
Swain Hall West 117, Bloomington IN 47405, USA
\newline
% ...
$^{ 16}$Department of Physics, Schuster Laboratory, The University,
Manchester M13 9PL, UK
\newline
% ...
$^{ 18}$Laboratoire de Physique Nucl\'eaire, Universit\'e de Montr\'eal,
Montr\'eal, Quebec H3C 3J7, Canada
\newline
% ...
$^{ 23}$Department of Physics and Astronomy, Tel Aviv University,
Tel Aviv 69978, Israel
\newline
$^{ 24}$International Centre for Elementary Particle Physics and
Department of Physics, University of Tokyo, Tokyo 113, and
Kobe University, Kobe 657, Japan
\newline
% ...
$^{ 29}$University of British Columbia, Department of Physics,
Vancouver BC V6T 1Z1, Canada
\newline
% ...
$^{  a}$ and at TRIUMF, Vancouver, Canada V6T 2A3
\newline
% ...

\newpage

\section{Introduction}
\label{sec-intro}

A large number of studies of inclusive vector-meson 
production in multihadronic Z$^0$ 
decay have so far concentrated on the measurement of fragmentation
functions and total inclusive rates (see for example~\cite{review} 
and~\cite{compilation} for reviews and data compilations). Relatively little 
has been done to investigate the role of meson spin in
the production dynamics. The primary quarks from Z$^0$ decay
at LEP are highly polarized, and this may play a role in
determining the helicities of leading vector mesons.
Measurements of  
B$^*$ mesons~\cite{AB*, DB*, OB*, Ospin} have shown that these 
are produced equally in all three helicity states, as would be expected
in a simple statistical picture~\cite{statistical, Bigi} where sea quark 
helicities are chosen at random and the ratio of vector to 
pseudoscalar meson production
is 3:1. However, in production of D$^*$ and $\phi(1020)$ mesons at 
large momentum fraction \xp~$(\equiv~p_{\mathrm meson}/p_{\mathrm beam})$, 
where the mesons have a high probability to contain a primary quark, 
OPAL have reported~\cite{Ospin} deviations from equal
populations of the three helicity states. The helicity density matrix
element \Rzz\ was found to be greater than $1/3$, corresponding to the 
helicity-zero state
being favoured. At the same time, the data showed evidence 
for negative values of the matrix
element $\Re \rh{1}{-1}$ for both D$^*$ and $\phi(1020)$,
in agreement with qualitative predictions of a model based on coherent 
(non-independent) fragmentation
of the primary quark and antiquark~\cite{coherent}.
Recently, DELPHI~\cite{Dspin} have reported measurements
for \rzz, \KS\ and $\phi(1020)$, confirming the OPAL observations of 
vector-meson spin
alignment at large \xp, but finding no evidence for non-zero values 
of the off-diagonal element $\Re \rh{1}{-1}$.
In the present paper, the OPAL studies are extended to the 
\KS\ meson (and its antiparticle) in its K$^+\pi^-$ (K$^-\pi^+$)
decay mode, and  
the helicity density matrix elements are measured
with high statistics over the entire \xp\ range.   
The findings are compared with a new theory~\cite{newAnselmino}
based on a Standard Model description of the spin structure of
the reaction \eetoZtoqq, followed by coherent
fragmentation of the \qqbar\ pair. 

\section{Formalism for the helicity density matrix}
\label{sec-formalism}

The formalism of the helicity density matrix
and its measurement using the vector-meson decay angular 
distribution
have been described fully in~\cite{Ospin}. The analysis is done 
in the vector-meson helicity rest frame, where the $z$-axis is 
the direction of the \ks\ in the overall centre-of-mass system 
(the same as the 
laboratory at LEP), the $y$-axis is the vector product of this $z$-axis
with the incident e$^-$ beam direction, and the $x$-axis is such as to
form a right-handed coordinate system. The angles 
$\theta_{\mathrm H}$ 
and $\phi_{\mathrm H}$ are the usual polar and azimuthal angles 
of the K$^\pm$ from the \ks\ decay, measured in this frame. 
The helicity density matrix elements, 
$\rho_{\lambda \lambda'}$, are determined by fitting 
the observed angular distributions $W$ using:
\begin{equation}
 W(\cos\theta_{\mathrm H}) = \frac{3}{4}~\left[(1-\rh{0}{0}) +
 (3\rh{0}{0}-1)\cos^{2}\theta_{\mathrm H}\right]
\label{thetadist}
\end{equation}
\begin{equation}
 W(|\alpha|) = (2/\pi)~[1 + 2\, \Re \rh{1}{-1} \cos2|\alpha|\ ]
\label{alphadist}
\end{equation} 
\begin{equation}
 W(|\beta|) = (2/\pi)~[1 + 2\, \Im \rh{1}{-1} \cos2|\beta|\ ]
\label{betadist}
\end{equation} 
The angles $\alpha$ and $\beta$ are introduced to exploit the
symmetry properties of the distributions and, with $\phi$ defined
over the range $-\pi$ to $\pi$, are given
by: $\alpha = |\phi_{\mathrm H}| - \pi/2$;
$\beta = |\phi_{\mathrm H} + \pi/4| - \pi/2$ for 
$\phi_{\mathrm H} < 3\pi/4$; and
$\beta = |\phi_{\mathrm H} - 3\pi/4| - \pi/2$ for 
$\phi_{\mathrm H} > 3\pi/4$. 
Two measured asymmetries give other combinations of matrix elements: 
\begin{eqnarray}
\Re(\rh{1}{0} - \rh{0}{-1}) 
& = 
& \!\!- \frac{\pi}{2 \sqrt{2}} \:\:
  \frac{N(\sin 2\theta_{\mathrm H} \cos\phi_{\mathrm H} > 0) 
 - N(\sin 2\theta_{\mathrm H} \cos\phi_{\mathrm H} < 0) } 
       {N(\sin 2\theta_{\mathrm H} \cos\phi_{\mathrm H} > 0) 
 + N(\sin 2\theta_{\mathrm H} \cos\phi_{\mathrm H} < 0) }
\label{re10}
% \\
\end{eqnarray}
\begin{eqnarray}
\Im(\rh{1}{0} - \rh{0}{-1}) 
& =
& \:\: \frac{\pi}{2 \sqrt{2}} \:\: 
 \frac{N(\sin 2\theta_{\mathrm H} \sin\phi_{\mathrm H} > 0 ) - 
           N(\sin 2\theta_{\mathrm H} \sin\phi_{\mathrm H} < 0) }
      {N(\sin 2\theta_{\mathrm H} \sin\phi_{\mathrm H} > 0 ) +  
           N(\sin 2\theta_{\mathrm H} \sin\phi_{\mathrm H} < 0) }
\label{im10}
\end{eqnarray}
where $N$ is the number of events in the given angular range.

The element \rh{0}{0}\  of the helicity density matrix gives the relative
intensity of mesons in the helicity 0 state, while the off-diagonal
element \rh{1}{-1}\ is a measure of coherence between the 
helicity $+1$ and helicity $-1$ states. 

\section{The OPAL detector and data samples}
\label{sec-OPAL}

The OPAL detector is described in~\cite{OPALdet}. For the 
present analysis, the most important components were the central
tracking chambers which consist of two layers of
silicon microvertex detectors~\cite{silicon}, a high-precision 
vertex drift chamber,
a large-volume jet chamber, and a set of drift chambers (the $z$-chambers)
which measure the coordinates of tracks along the direction of
the beam. The OPAL coordinate
system is defined with the $z$-axis following the electron beam direction; 
the polar angle~$\theta$~is
defined relative to this axis, and $r$ and $\phi$ are the usual
cylindrical polar coordinates.
The central chambers lie within a homogeneous axial magnetic field
of $0.435$~T. Charged particle tracking is
possible over the range $|\cos \theta|<0.98$ for the full range of 
azimuthal angles. 
For K$\pi$ systems around the \ks\ mass, the
mass resolution of the detector varies, with momentum, from about 5 to 20~MeV.
The OPAL jet chamber is capable of measuring specific energy loss, \dedx,
with a resolution, $\sigma(\dedx)/(\dedx)$, of 3.5\% for 
well-reconstructed, high-momentum 
tracks in multihadronic events~\cite{DEDX}.

The present analysis used the full OPAL sample of 4.3 million multihadronic 
Z$^0$ decays recorded at LEP 1 between 1990 and 1995. To correct for 
losses due to the acceptance and efficiency of the experiment 
and the selection procedures, and also to provide signal and 
background shapes for fits to the data mass spectra, 6 million 
Monte Carlo events were used, which had been generated using
JETSET 7.4~\cite{JETSET} tuned to the OPAL data~\cite{JTtune}, 
and processed
through a full simulation of the experiment~\cite{GOPAL} and the 
data reconstruction and analysis. 

A detailed description of the selection of hadronic Z$^0$ decay events 
in OPAL is given in~\cite{TKMH}. For the present analysis, tracks 
in the selected events were
required to have: a minimum momentum transverse to the beam direction
of 150~MeV/$c$; a maximum momentum of $1.07 \times E_{\mathrm {beam}}$,
based on the the momentum resolution of the detector; a distance
of closest approach to the interaction point less than 5~cm in the 
plane orthogonal
to the beam direction, and the corresponding
distance along the beam direction less than 40~cm; 
a first measured point within a radius of 75~cm 
from the vertex; and at least 20 hits available for measurement of
specific energy loss, \dedx.
 
Kaons and pions were identified using the \dedx\ measurements.
For each track, a $\chi^2$ probability (weight) was formed for each of the
stable particle hypotheses (e, $\mu$, $\pi$, K and p). A track was 
identified as a pion or a kaon if
the appropriate weight was above 5\% and was larger than the weight for
each of the other stable hadron hypotheses. 
Between momenta of 0.8 and 2.0~GeV/$c$, 
the $\pi$, K and p bands overlap in \dedx, leading to considerable
ambiguity among hypotheses. Therefore no tracks were identified as kaons 
in this momentum range although, since most tracks are pions, 
pion identification was still allowed. With these \dedx\ selections,
pions were identified with a typical efficiency of 50\% and a sample 
purity of 90\% over most of the momentum range. The kaon efficiency was
around 45\%, while the sample purity was typically 30\% at low momentum,
rising to 50\% at high momentum.

\section{Measurement of the 
\mbox{\boldmath ${\mathrm K}^*(892)^0$}
helicity density matrix and 
fragmentation function}
\label{sec-measurement}

\subsection{Inclusive 
\mbox{\boldmath ${\mathrm K}^\pm \pi^\mp$} mass spectra}

For both the real and simulated data samples,
inclusive mass spectra were formed for \Kpi\ and \likesign\ systems
in ranges of scaled momentum, \xp, of the 2-particle systems. Within
each of 12 \xp\ ranges, whose limits are given in table~\ref{results}, 
the data were further divided, based on
the K$^\pm$ direction in the K$\pi$ helicity frame, into each of six bins of 
\CTH, four bins of $|\alpha|$, four bins of $|\beta|$ and, for asymmetry 
measurements, the positive and negative regions of each of 
$\sin 2\theta_{\mathrm H}\cos \phi_{\mathrm H}$
and $\sin 2\theta_{\mathrm H}\sin \phi_{\mathrm H}$.
This resulted in $12 \times 18 = 216$ mass spectra for each 
of \Kpi\ and \likesign.
To account for combinatorial backgrounds, some of which change
rapidly near the \ks\ peak region due to kinematics, background-subtracted
mass spectra were formed by subtracting the like-charge \likesign\
spectra from those for \Kpi. For the Monte Carlo sample, separate
spectra were also made for the most important states contributing
to \Kpi, using information on the origin of each track
at the generator level.

\begin{figure}[htbp]
\begin{center}
\mbox{\epsfig{file=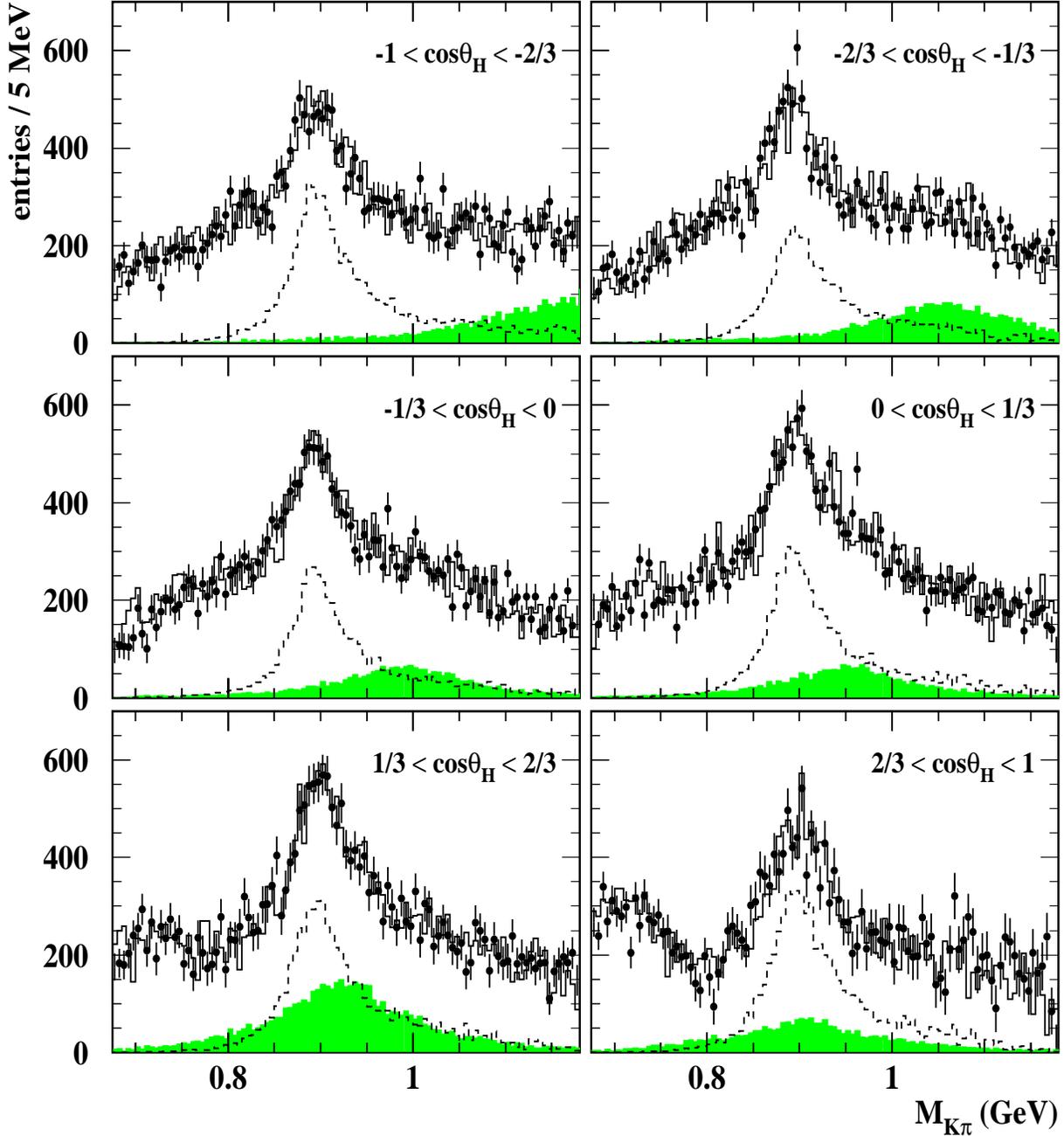,height=20cm,width=18cm}}
\end{center}
\caption{Inclusive background-subtracted \Kpi\ mass spectra for
the \xp\ range $0.3 < x_p < 0.4$ for the 6 bins of \CTH. The points with
error bars show the OPAL data.
The full histograms show the fits described in section 4.3, with the dashed 
histograms giving the \ks\ component and the shaded areas 
showing the contributions of the \rz\ reflection.}
\label{fig-mass}
\end{figure}

Imperfect particle identification leads to significant contributions
in the \Kpi\ mass spectra from \pipi. The \rzz\ reflection is
particularly important. For kinematical reasons, its shape and 
peak position vary with \xp\ and with the reconstructed 
\CTH\ of the misidentified pion. In general, the \rz\ reflection 
peaks well above the mass of the \ks\ for negative values of 
\CTH\ but moves under the \ks\ peak as \CTH\ approaches +1. 
Figure~\ref{fig-mass}, which shows the background-subtracted
\Kpi\ mass spectra 
in the six \CTH\ bins for $0.3<x_p<0.4$, illustrates this behaviour 
of the \rz\ reflection. The \pipi\ products of the \om\ decay 
also peak close to the \ks\ and move with \CTH, but they form a
broader distribution. Other states, such as 
$\phi(1020) \rightarrow {\rm K}^+{\rm K}^-$, are less important and are
negligible near to the \ks\ peak when compared to the 
combinatorial backgrounds.

\subsection{Efficiency for \mbox{\boldmath ${\mathrm K}^{*0}$} reconstruction}
\label{efficiency}

The overall reconstruction efficiency for \ks\ decaying via \Kpi\ 
was measured separately in each of the 216
bins (of \xp\ plus an angular variable) using the Monte Carlo simulation. 
The efficiency typically 
varied between 10\% and 25\% with
\CTH\ for all \xp\ above 0.1, while at lower \xp\ it 
was a stronger function of \CTH, varying from 5\% to 35\%.
The efficiency depended much less on the azimuthal angles, $\alpha$ 
and $\beta$. 

Due to the track selection and particle identification criteria,
and particularly to the fact that no kaons were identified in the
overlap region of \dedx\ from 0.8 to 2~GeV/$c$,
the \ks\ efficiency was zero for some regions in the 
\xp\ versus \CTH\ plane, particularly for $x_p<0.15$. The 
limits of the \xp\ bins, listed in table~\ref{results}, were chosen 
to ensure that each bin of \xp\ contained a measurable \ks\ signal for every
bin of \CTH. Because the decay particles have a momentum of
about 300~MeV/$c$ for a \ks\ at rest, measurements were possible 
down to $x_p=0.$   

\subsection{Measurements of the helicity density matrix}
\label{sec-fits}

Each of the 216 background-subtracted mass spectra was fitted using 
a minimum $\chi^2$ fit, for 100 bins of mass over
the range $0.675-1.175$~GeV, to a 
sum of contributions (as histograms) from \KS\ signal, 
\rzz\ reflection, \omm\ reflection and background. For the \ks\ signal, 
the shape was
taken from the detector-level Monte Carlo simulation, but
because the \ks\ is generated in JETSET as a simple Breit-Wigner
resonance, the output of the simulation was reweighted as a 
function of mass to reproduce a P-wave Breit-Wigner with mass-dependent 
width, which is expected
to represent the \ks\ line shape in the real data. This procedure
automatically takes account of variations of mass resolution with
momentum and decay angles. The shapes of the \om\ reflections 
were also taken from the simulation. To obtain the shapes of 
the residual combinatorial backgrounds for the fits, the Monte
Carlo spectra were treated in the same way as the data, and then the
contributions from \ks\ and the \rz\ and \om\ reflections were
removed. 

The shape of the \rzz\ resonance in Z$^0$ decay has long been known
to be severely distorted from a Breit-Wigner shape~\cite{OPALrho},
particularly at low \xp\ where the peak mass is shifted by some 
40~MeV. The source of the distortion is still uncertain, but is likely to 
be due to residual Bose-Einstein correlations~\cite{RBEC}. 
The standard implementation of Bose-Einstein correlations in the 
JETSET program has been found to improve significantly the agreement with 
data~\cite{OPALrho, DELrho}. The ALEPH Collaboration~\cite{Arho} have 
measured inclusive \rz\ production by allowing, in their \pipi\ mass 
spectrum fits, variation of the JETSET parameters which control the 
strength of the correlations. As the standard representation of the \rz\ 
shape for the present analysis, the fitted ALEPH values have been used to 
generate \rz\ line shapes using JETSET. Since the Bose-Einstein effects act 
primarily on low-momentum particles, shapes were generated as  
functions of both \xp\ and \CTH\ since for a given value of \xp\ 
the momentum of a decay pion is related to \CTH. 
As a result the line shape of the \rz, and consequently its reflection in 
\Kpi, varies with both \xp\ and \CTH. To obtain the shapes of
the reflections in \Kpi, the decay tracks from the generated
\rz\ mesons were passed through
a Monte Carlo model with a simplified simulation of the detector and
the analysis selections. The performance of this simulation was verified by 
comparing its results, for a simple Breit-Wigner shape as input, with 
the output of the full detector-level JETSET simulation. 
Other possible \rz\ line shapes were considered for studies of
systematic errors, and will be discussed in section~\ref{sec-rhoshape}.

Figure~\ref{fig-mass} shows, as an example, the results of the 
mass spectrum fits 
for the six bins of \CTH\ in the \xp\ bin $0.3 < x_p < 0.4$. 
The average value of $\chi^2$ for the fits, all of which were
acceptable, to the 72 independent
% 120 independent mass spectra (in the \CTH\ and $\alpha$ bins)
% was 95.4 for 96 degrees of freedom.
mass spectra in bins of \CTH\ and \xp\ was 99.2 for 96 degrees of freedom.
The \ks\ intensities extracted from the fits were
fully corrected for efficiency and branching ratio to give, in each of 
the 12 bins of \xp, the three inclusive double differential cross 
sections for \ks\ production and decay, 
$(1/\sigma_{\mathrm h}){\mathrm d}^2\sigma/{\mathrm d}{x_p}{\mathrm d}y$,
for $y = \cos \theta_{\mathrm H}$, $|\alpha|$ and  $|\beta|$, where
$\sigma_{\mathrm h}$ is the total multihadronic Z$^0$ cross section.

\begin{figure}[htbp]
\begin{center}
\mbox{\epsfig{file=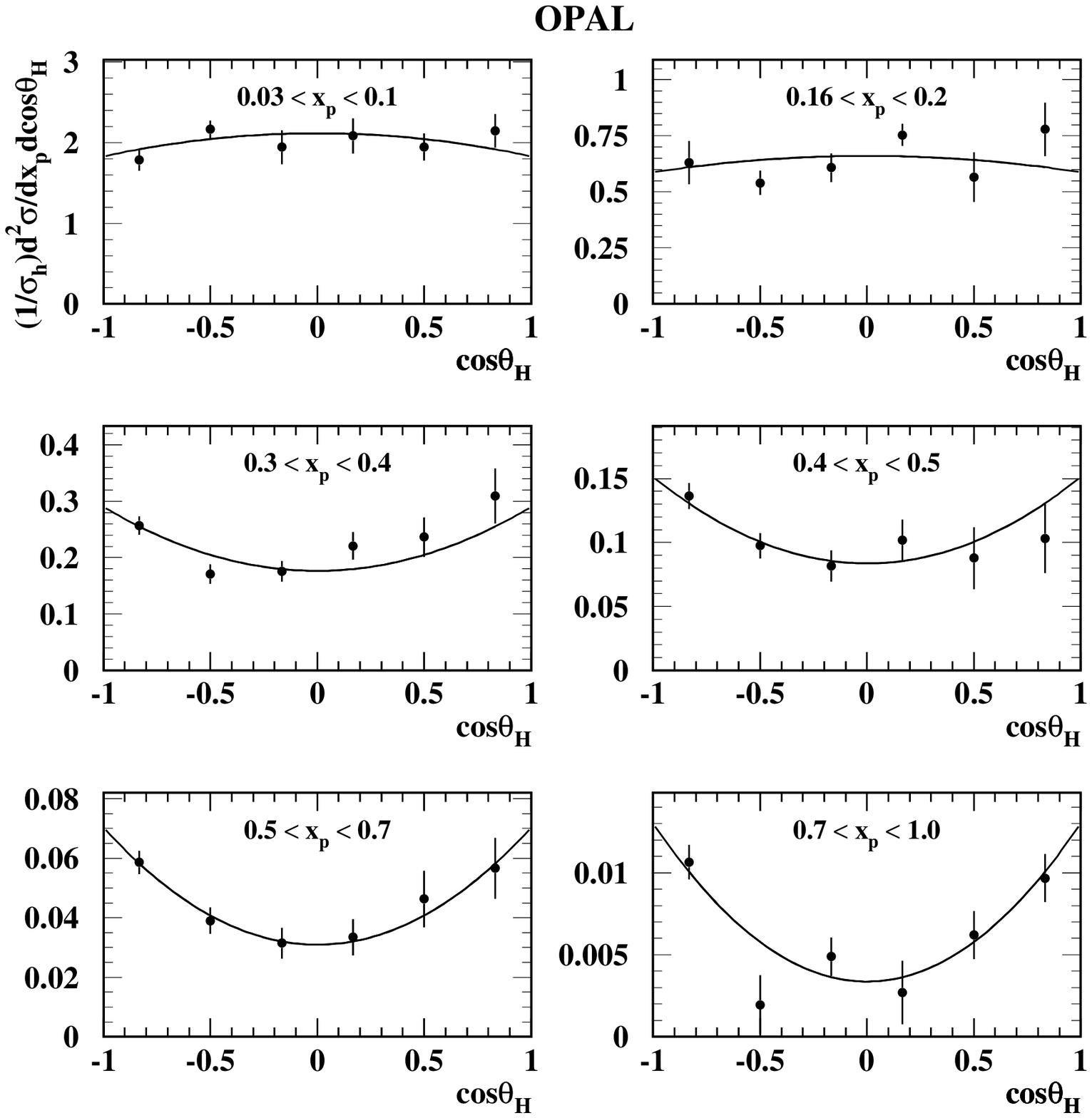,height=20cm,width=18cm}}
\end{center}
\caption{Differential cross sections, \ddiffc, for 6 ranges of \xp.
The points are the OPAL data and
the curves show the fits used to obtain the helicity density matrix
elements \Rzz. The errors are statistical only.}
\label{fig-costh}
\end{figure}

\begin{figure}[htbp]
\begin{center}
\mbox{\epsfig{file=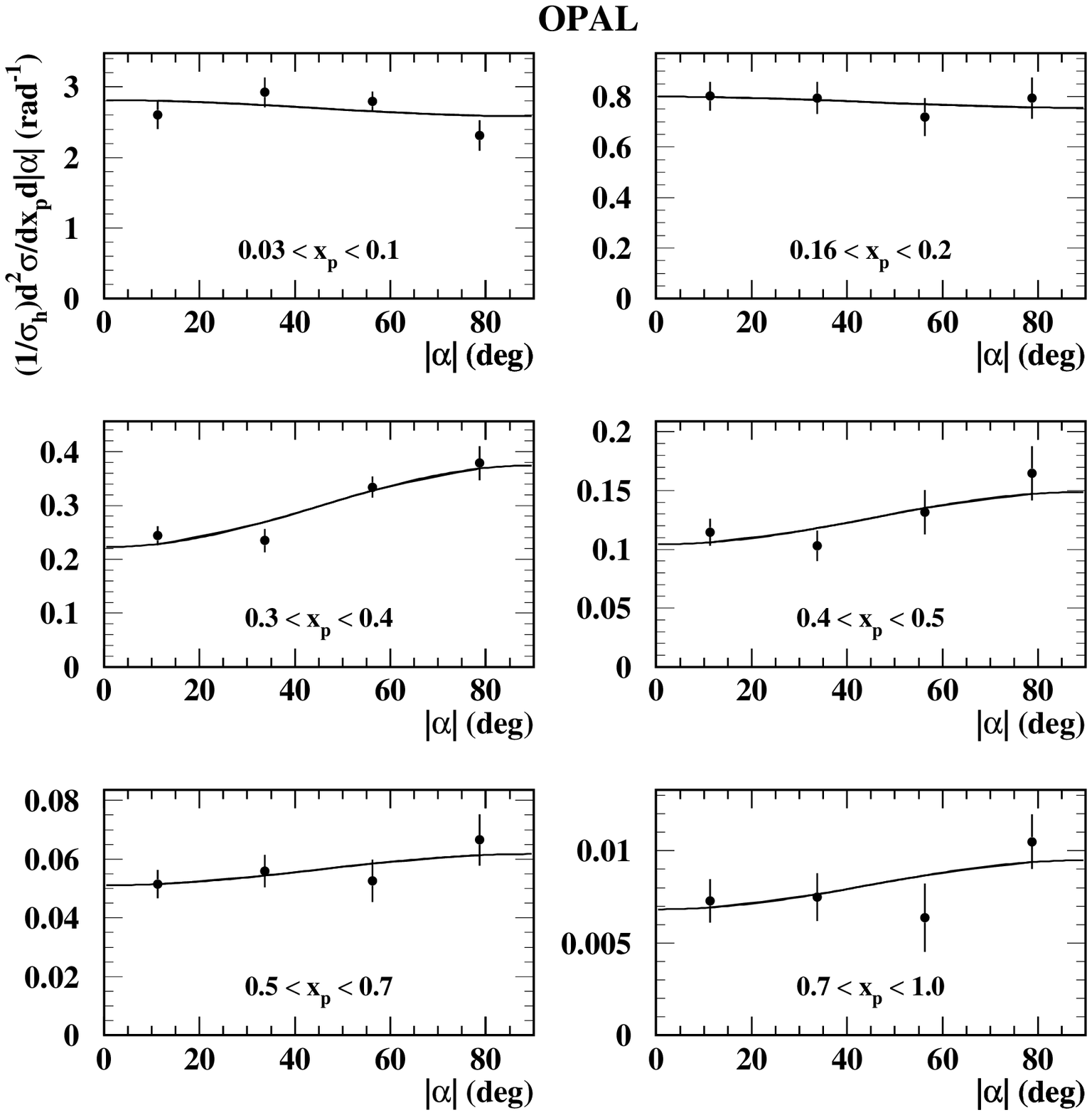,height=20cm,width=18cm}}
\end{center}
\caption{Differential cross sections, \ddiffa, for 6 ranges of \xp.
The points are the OPAL data and
the curves show the fits used to obtain the helicity density matrix
elements \Rer. The errors are statistical only.}
\label{fig-alpha}
\end{figure}

The helicity density matrix elements in each \xp\ bin were  
obtained by fitting these cross sections to the appropriate form 
%(equations~(\ref{thetadist}), (\ref~{alphadist}) and~(\ref{betadist})) 
(equations~(1), (2) and~(3)) 
as functions of the angular variables. Figures~\ref{fig-costh} 
and~\ref{fig-alpha} show some of the
measured differential cross sections in \CTH\ and $|\alpha|$
along with the results of the fits.
The two asymmetries, in 
$\sin 2\theta_{\mathrm H}\cos \phi_{\mathrm H}$ 
and $\sin 2\theta_{\mathrm H}\sin \phi_{\mathrm H}$, were also measured
(equations~(\ref{re10}) and~(\ref{im10})).

As outlined in section~\ref{efficiency}
the \ks\ efficiencies were measured for bins of \CTH\ and \PHIH, each 
integrated over the range of the other, using Monte Carlo events which 
were generated with isotropic decay distributions. Since the data clearly
show non-isotropic decay distributions at large \xp, any correlations
in efficiency between \CTH\ and \PHIH\ could therefore bias the 
results of the measurements. The fits were therefore iterated, at
each stage using the fitted values of the density matrix elements
to generate angular distributions with which to recalculate the 
efficiencies in \CTH\ and \PHIH\ for the next iteration. The procedure
produced only small changes, well within the statistical errors, to the
measurements. The final results quoted are those obtained for two iterations,
after which no further changes were induced within the quoted number 
of significant figures. 

\begin{figure}[htbp]
\begin{center}
\mbox{\epsfig{file=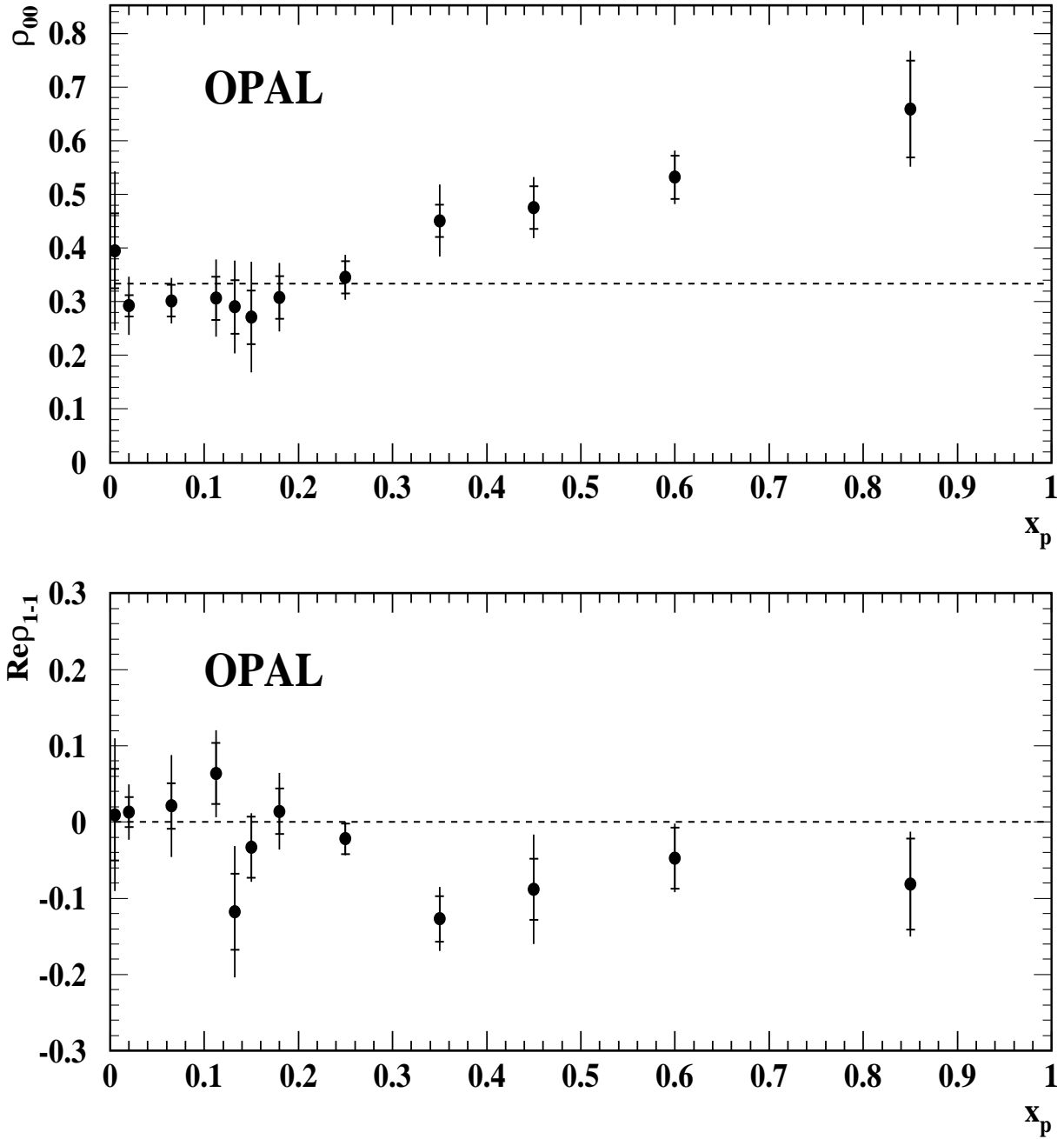,height=20cm,width=18cm}}
\end{center}
\caption{Measured \ks\ helicity density matrix elements, \Rzz\ and \Rer,
as functions of \xp. The error bars are statistical and systematic combined
in quadrature, with the tick marks giving the statistical errors only.}
\label{fig-sdmes}
\end{figure}

The measured values of $\Im \rh{1}{-1}$, $\Re(\rh{1}{0} - \rh{0}{-1})$ and
$\Im(\rh{1}{0} - \rh{0}{-1})$ were found, as expected, to be consistent with
zero over the entire $x_p$ range, with weighted average values for
$x_p>0.3$ of $-0.03\pm0.03$, $-0.06\pm0.05$ and $-0.02\pm0.04$ 
respectively.  
For $x_p < 0.3$
the values of \Rzz\ and \Rer\ were also consistent with an
isotropic decay distribution. However
in the large momentum region, $x_p > 0.3$, the
diagonal element \Rzz\ was found to be larger than 1/3 while
\Rer\ was negative. Table~\ref{results} gives
the measured values of \Rzz\ and \Rer\ in the bins of \xp. 
Systematic errors, given in the table, are discussed below in
section~\ref{sec-systematics}.
The data are also shown in figure~\ref{fig-sdmes} where the clear
deviations from \Rzz$=1/3$ and \Rer$=0$ are
seen for $x_p > 0.3$, indicative of a
`leading particle effect', i.e. the probability for the meson to contain
one of the primary quarks rises as the scaled momentum 
increases; this is discussed further in section 6.

\subsection{Fragmentation function and total rate}
\label{sec-fragf}

For each of the 12 \xp\ bins, the fragmentation function, 
$(1/\sigma_{\mathrm h}){\mathrm d}\sigma/{\mathrm d}x_p$, 
was measured by integrating the fitted curves obtained from 
the \CTH\ spectra. The results are given in table~\ref{results}. 
The total inclusive rate for \ks\ production was obtained by
integrating the fragmentation function. Since the
differential cross section has been measured over the entire \xp\
range, no explicit systematic uncertainty arises from extrapolation 
using a Monte Carlo model. However there is an implicit assumption,
borne out by the measurements, that the angular distribution in the
Monte Carlo model is the same as that for the data for
$0.03<x_p<0.1$ where there is a significant region of zero 
efficiency in the \CTH\ versus \xp\ plane. The overall inclusive
rate was determined to be $0.74 \pm 0.02 \pm 0.02$ \KS\ mesons per
multihadronic Z$^0$ decay, in agreement with previous 
measurements~\cite{OK*, DK*, Arho} (but with smaller
statistical and systematic errors). These results
update the previous OPAL measurements~\cite{OK*} of the 
\KS\ fragmentation function and total inclusive rate. Since the
present measurements have been obtained by
integrating the cross section,
$(1/\sigma_{\mathrm h}){\mathrm d}^2\sigma/
{\mathrm d}{x_p}{\mathrm d}\cos \theta_{\mathrm H}$, they are
in principle more accurate at large $x_p$ than all previous measurements;
in previous analyses the data were not binned in 
$\cos \theta_{\mathrm H}$,
and efficiencies were calculated assuming isotropic 
decay distributions. 

\section{Systematic effects}
\label{sec-systematics}

\subsection{The \mbox{\boldmath $\rho(770)^0$} line shape}
\label{sec-rhoshape}

An important source of systematic error on the \ks\ measurements 
clearly arises from uncertainty in the shape of the \rzz\ resonance and its
reflection in the \Kpi\ mass spectrum. To measure the systematic error 
from this source, the fits were repeated using two extreme possibilities 
for the shapes: those obtained from the simple Breit-Wigner resonance
in the full detector-level simulation; and shapes obtained by 
multiplying a relativistic P-wave Breit-Wigner with a skewing 
factor which varied with \xp, as used in~\cite{RBEC, OK*}.
The former shape has no skewing of the line shape and is known to give 
poor fits to the \pipi\ mass
spectra, particularly at low \xp. The latter includes a shift of the
peak position and a skewing of the line shape to low masses, but produces 
too sharp a cut-off at high mass. 
Because the \rz\ reflection is more important at \CTH~$ > 0$, the
entire analysis was also repeated using only the negative region of \CTH. 
For each of the measurements of 
inclusive cross sections and helicity density matrix elements 
the systematic error from the \rz\ lineshape uncertainty
was taken to be $1/\sqrt{12}$
times twice the maximum deviation obtained from the standard fit value.

\subsection{Detector and \mbox{\boldmath ${\rm d}E/{\rm d}x$} simulation}
\label{sec-systdetector}

The analysis was repeated, dividing the data into two approximately 
equal samples covering
two ranges of \ks\ production angle: 
$|\cos\theta| < 0.5$ and $|\cos \theta| > 0.5$. 
Here $\theta$ is the angle between the \Kpi\ system and the
incident electron beam. The 
former range is entirely in the barrel region of OPAL, while the latter
covers part of the barrel plus the whole of the end cap regions. 
In the barrel region, tracks may have the maximum number of jet
chamber hits; at higher values of $|\cos \theta|$
the number of possible hits falls and the systematics of \dedx\ measurements
change due to the low angle of the tracks with respect to the wires
of the jet chamber.
As well as serving as a check for any systematic differences arising from
the detector simulation of these two regions, this analysis provided 
results for comparison with theoretical predictions (see
section~\ref{sec-theory}). The measurements of the cross sections
and \Rzz\ values were in good agreement for the two regions, 
within the statistical errors, and so no systematic errors were assigned
from this study. 
The values of \Rer\ may be expected 
to vary with $\cos \theta$, as will be discussed in 
section~\ref{sec-theory}. 

To investigate further the effects of the \dedx\ requirements, the minimum 
number of hits required on a track was doubled to 40, and the analysis
repeated, with systematic errors taken to be the
maximum deviation of any measurement from its standard fit value.

Further systematic checks on the simulation of the energy loss were
made by varying the assumed mean values of the theoretical \dedx\ 
distributions for a given track hypothesis, and the assumed resolution on the
energy loss measurements. Studies of well-identified
pions from K$^0_{\rm S}$ decays, protons from $\Lambda$ decays and
kaons from D$^0$ and $\tau$ decays were used to place limits on the
maximum possible deviations of these quantities, and the analysis
was repeated, with the \dedx\ weights of the tracks being recalculated 
each time. Systematic errors were again assigned as $1/\sqrt{12}$ 
times twice the 
maximum measured deviations from the standard fit values.

\subsection{Influence of charm decays}

Decays of charmed particles of the type 
${\mathrm P} \rightarrow {\mathrm K}^{*0} + {\mathrm P}'$,
where P and ${\mathrm P}'$ denote any pseudoscalar mesons 
(for example $\overline {{\mathrm D}^0} \to $K$^{*0}\pi^0$),
must produce \ks\ mesons 
which are in the helicity-zero state when viewed from the
P rest frame. The total contribution of all 
such known decays~\cite{PDG} to the overall \ks\ spin 
alignment in the helicity frame has been studied using 
Monte Carlo and was found to be negligible. Their removal would
have no measurable effect below $x_p = 0.2$ and would
reduce the measured values of \rh{0}{0} by a maximum of 0.02 at 
$x_p$ around 0.5; their effect was therefore ignored. 

\subsection{The K$^{*0}$ line shape}

To take account of possible long tails in the 
line shape of the \ks\ resonance beyond the upper limit of the fits
to the mass spectra, the relativistic P-wave Breit-Wigner
used for the fits was integrated out to ten full widths above the
nominal peak position, resulting in an increase of $5\%$ in the total
intensity. Since the shape of such a resonance is uncertain
so far from the pole position, one half of this extra contribution 
was added to each measurement of the differential cross 
section (the results in table 1 already include these
corrections), and an additional systematic error of $5\%/\sqrt {12}$
was assigned to each measurement and to the total integrated rate.
This systematic error has no influence on the measurements of the
spin density matrix elements. 

The final systematic errors, as given in table~\ref{results}, were
obtained by combining in quadrature the contributions from all of
the above sources. 
For all of the measurements (of cross sections and spin density matrix
elements) the systematic errors from
detector and \dedx\ effects were found to be approximately equal to 
those from the \rz\ line shape uncertainties. 

\section{Comparison of results with theory}
\label{sec-theory}

In a recent paper~\cite{newAnselmino}, a relation is derived 
between the values of \Rzz\ and \Rer\ for vector mesons containing
a primary quark, using Standard Model 
parameters and some plausible assumptions about hadronization.
According to this theory, negative values of the off-diagonal 
element \Rer\ are 
generated for leading vector mesons by coherence in the
production and  
fragmentation of the primary quark and antiquark from the Z$^0$ 
decay. For the primary \ks\ mesons, it is 
predicted that
\begin{equation}
{\mathrm {Re}}~\rho_{1-1} \approx -0.17 (1-\rho_{00})
{\sin^2\theta \over (1+\cos^2 \theta)}
\label{eqn-ratio}
\end{equation}
where $\theta$ is the production angle of the leading 
\ks\ relative to the 
electron beam. In the present analysis, there is no explicit cut 
on $|\cos \theta|$, although the efficiency falls rapidly for
values above 0.9. Based on the distribution of high-\xp\ \ks\ production
angles in the detector-level Monte Carlo events, the ratio
${\mathrm {Re}}\rho_{1-1}/(1-\rho_{00})$ is 
predicted by equation~(\ref{eqn-ratio}) to be approximately $-0.10$ for
mesons containing a primary quark.
Table~\ref{ratios} shows the measured values at large \xp, with
statistical and systematic errors combined. The 
results in each independent \xp\ bin are compatible with the theory, 
although there is a suggestion that the magnitudes of the measured 
values are larger than predicted. The weighted average value is
$-0.19\pm0.05$ for $x_p > 0.3$. It should be noted however that
comparison with the theory is somewhat obscured by lack of knowledge
of the proportion of \ks\ mesons which contain a primary quark
at a given value of \xp. Studies using the JETSET model suggest that
this fraction is around $20\%$ at $x_p = 0.3$, rising approximately
linearly to $100\%$ at $x_p = 1.$

As discussed in section~\ref{sec-systdetector}
the density matrix elements were measured separately 
for two ranges of \ks\ production angle: 
$|\cos\theta| < 0.5$ 
and $|\cos \theta| > 0.5$. The results are given in table~\ref{thetabands}. 
The dependence of \Rer\ on $\theta$ predicted by equation~(\ref{eqn-ratio})
would imply that the ratio of these two 
measurements should be 2.3, assuming no variation of \Rzz\ with
$\cos\theta$. The measured ratio of the weighted mean values 
is $1.5 \pm 0.7$ over the range $x_p > 0.3$, consistent within 
relatively large errors both 
with the theory and with no variation of \Rer\ with $\cos\theta$.

While the theory of~\cite{newAnselmino} derives a relationship 
between the values of \Rzz\ and \Rer, it does not offer any explanation 
for deviations of \Rzz\ from 1/3. A number of models relevant to this
topic are discussed in~\cite{Ospin} and~\cite{Dspin}. 
In the simple statistical model
based on spin counting~\cite{statistical}, the maximum value of \Rzz\ 
is 0.5, and this can only occur when there is zero probability for 
production of pseudoscalar mesons containing the primary quark. 
While this model fits with the measurements of 
B$^*$ mesons~\cite{AB*, DB*, OB*, Ospin},
it is now clearly ruled out by the LEP data for the lighter 
vector mesons. The QCD-inspired model
of~\cite{Augustin}, which predicts \Rzz$=0$ for leading vector mesons
is also ruled out by the measurements; however~\cite{Augustin} does
point out that this result would only be expected in the limit that
quark and meson masses and transverse momenta can be neglected. The model 
of~\cite{statistical},
in which vector-meson production is considered as arising from
the helicity-conserving process q$\rightarrow$qV, predicts \Rzz$=1$
for the leading vector mesons. Although its result is  
in accord with the observations, this model is based on a picture of 
parton hadronization which has been superseded by the more firmly based,
detailed and successful string and cluster models; these latter models do not
however address the dynamical role of meson spin in the hadronization.

\section{Conclusions}
\label{sec-conclusions}

From the present analysis, together with OPAL measurements~\cite{Ospin} 
of inclusive $\phi$ and 
D$^*$ mesons, and DELPHI~\cite{Dspin} measurements of
$\rho^0$, K$^{*0}$ and $\phi$, it is now clearly established
that leading vector mesons, other than the B$^*$, are 
spin-aligned with a preference for the helicity-zero state. 
% There are currently no theories which explain these observations.

The results of the present \ks\ analysis are consistent with the idea
that coherence in the \qqbar\ production and fragmentation plays a 
role in generating 
non-zero values of the off-diagonal element \Rer, in agreement
with the theory of~\cite{newAnselmino}. 
However the measured value of \Rer\ for the K$^{*0}$ is slightly larger
than the theory predicts, while for the D* it is smaller~\cite{Ospin}.
This may indicate a flavour dependence of the mechanism producing
the spin alignment, or the existence of more than one mechanism.
While OPAL observe
non-zero (and negative) values of \Rer\ in three separate analyses,
of K$^{*0}$, $\phi$ and D$^*$ production, the DELPHI results on
$\rho^0$, \ks\ and $\phi$ are all consistent with zero, within
relatively small errors. Further measurements are required to clear
up this apparent inconsistency.

The LUND string model~\cite{string}, as implemented in the 
JETSET Monte Carlo program~\cite{JETSET}, is highly successful in describing
many features of hadronic Z$^0$ decay events~\cite{toprev}. However, 
in the simplest string model, no spin alignment of vector mesons is
expected~\cite{string}. Similarly, the QCD cluster model of
HERWIG~\cite{cluster} has no mechanism to produce spin-aligned
vector mesons; and neither model will generate non-zero values for the
off-diagonal elements of the helicity density matrix. 
The vector meson spin alignment results
therefore provide a challenge to the standard Monte Carlo models
of the physics of inclusive hadron production in Z$^0$ decay at LEP. 
Can they be extended to include a physical mechanism which 
explains the measured vector-meson helicity density matrices? 

\bigskip
\bigskip
\noindent
{\bf Acknowledgements}
\par
We particularly wish to thank the SL Division for the efficient operation
of the LEP accelerator at all energies
 and for
their continuing close cooperation with
our experimental group.  We thank our colleagues from CEA, DAPNIA/SPP,
CE-Saclay for their efforts over the years on the time-of-flight and trigger
systems which we continue to use.  In addition to the support staff at our own
institutions we are pleased to acknowledge the  \\
Department of Energy, USA, \\
National Science Foundation, USA, \\
Particle Physics and Astronomy Research Council, UK, \\
Natural Sciences and Engineering Research Council, Canada, \\
Israel Science Foundation, administered by the Israel
Academy of Science and Humanities, \\
Minerva Gesellschaft, \\
Benoziyo Center for High Energy Physics,\\
Japanese Ministry of Education, Science and Culture (the
Monbusho) and a grant under the Monbusho International
Science Research Program,\\
German Israeli Bi-national Science Foundation (GIF), \\
Bundesministerium f\"ur Bildung, Wissenschaft,
Forschung und Technologie, Germany, \\
National Research Council of Canada, \\
Hungarian Foundation for Scientific Research, OTKA T-016660, 
T023793 and OTKA F-023259.\\

\vfill

\newpage
\begin{table}[htb]
\begin{center}
\begin{tabular}{|c|r@{--}l|r@{$\pm$}c@{$\pm$}l|r@{$\pm$}c@{$\pm$}l|r@{$\pm$}c@{$\pm$}l|}
\hline
\xp\ bin &\multicolumn{2}{c|}{\xp\ range} &  \multicolumn{3}{c|}{ \frag\ } &
\multicolumn{3}{c|}{ \Rzz\ } & \multicolumn{3}{c|}{ \Rer\ } \\
\hline
1 &0.0&0.01   & 1.22&0.15&0.04   & 0.39&0.07&0.13 &   0.01&0.06&0.08   \\
2 &0.01&0.03  & 4.96&0.17&0.15   & 0.29&0.02&0.05 &   0.01&0.02&0.03   \\
3 &0.03&0.1   & 4.14&0.20&0.19   & 0.30&0.03&0.03 &   0.02&0.03&0.06   \\
4 &0.1&0.125  &2.35&0.16&0.13    & 0.31&0.04&0.06 &   0.06&0.04&0.04   \\
5 &0.125&0.14 &1.99&0.15&0.09    & 0.29&0.05&0.07 & $-0.12$&0.05&0.07  \\
6 &0.14&0.16  &1.60&0.11&0.10    & 0.27&0.05&0.09 & $-0.03$&0.04&0.02  \\
7 &0.16&0.2   &1.30&0.09&0.06    & 0.31&0.04&0.05 &    0.01&0.03&0.04  \\
8 &0.2&0.3    &0.81&0.04&0.05    & 0.34&0.03&0.03 & $-0.02$&0.02&0.01  \\
9 &0.3&0.4    &0.44&0.03&0.03    & 0.45&0.03&0.06 & $-0.13$&0.03&0.03  \\
10&0.4&0.5    &0.22&0.02&0.01    & 0.48&0.04&0.04 & $-0.09$&0.04&0.06  \\
11&0.5&0.7    &0.090&0.009&0.003 & 0.53&0.04&0.03 & $-0.05$&0.04&0.02  \\
12&0.7&1.0    &0.013&0.004&0.003 & 0.66&0.09&0.06 & $-0.08$&0.06&0.03  \\
\hline 
\end{tabular}
\end{center}
\caption{Inclusive \ks\ cross sections and helicity density matrix elements
measured in the helicity-beam frame over the full range of \ks\ 
production angles. The first errors are statistical and the second
systematic.}
\label{results}
\end{table}

\begin{table}[htb]
\begin{center}
\begin{tabular}{|c|c|}
\hline
 $x_p$ range & \Rer/($1-$\Rzz) \\
\hline
0.3--0.4  &  $-0.24\pm 0.08$ \\
0.4--0.5  &  $-0.17\pm 0.13$ \\
0.5--0.7  &  $-0.11\pm 0.09$ \\
0.7--1.0  &  $-0.24\pm 0.22$ \\
\hline
0.3--1.0  &  $-0.19\pm 0.05$ \\
\hline
\end{tabular}
\end{center}
\caption{Ratios of \ks\ helicity density matrix elements
measured over the full range of production angles at large \xp.
The last row gives the weighted
mean value over the range indicated.}
\label{ratios}
\end{table}

\begin{table}[htb]
\begin{center}
\begin{tabular}{|c|c|c|}
\hline
 $x_p$ range & $|\cos \theta| < 0.5$ & $ |\cos \theta| > 0.5$ \\
\hline
0.3--0.4  &  $-0.12 \pm 0.04$ & $-0.13 \pm 0.04$ \\
0.4--0.5  &  $-0.14 \pm 0.06$ & $-0.05 \pm 0.06$ \\
0.5--0.7  &  $-0.08 \pm 0.05$ & $-0.02 \pm 0.06$ \\
0.7--1.0  &  $-0.17 \pm 0.10$ & $-0.01 \pm 0.11$ \\
\hline
0.3--1.0  &  $-0.12 \pm 0.03$ & $-0.08 \pm 0.03$ \\
\hline
\end{tabular}
\end{center}
\caption{Values of \Rer\ measured over different ranges of \ks\
production angles at large \xp. The last row gives the weighted
mean values over the range indicated. Statistical errors only are
given; the systematic errors largely cancel when the ratios are taken.}
\label{thetabands}
\end{table}

\newpage

\newpage

\newpage

\newpage

\end{document}